\documentclass[%
preprint,
 amsmath,amssymb,
 aps,
pra,
]{revtex4-2}

\usepackage{graphicx}
\usepackage{dcolumn}
\usepackage{bm}

\usepackage{tabularx}
\usepackage{algpseudocode}
\usepackage{algorithm}

\usepackage[utf8]{inputenc}
\usepackage[T1]{fontenc}
\usepackage{mathptmx}
\usepackage{etoolbox}
\usepackage{xcolor}

\begin{document}


\title{Unlocking Hidden Information in Sparse Small-Angle Neutron Scattering Measurements}

\author{Chi-Huan Tung$^1$}
\author{Sidney Yip$^2$}
\author{Guan-Rong Huang$^3$}
\author{Lionel Porcar$^4$}
\author{Yuya Shinohara$^5$}
\author{Bobby G. Sumpter$^6$}
\author{Lijie Ding$^1$}
\author{Changwoo Do$^1$}
\author{Wei-Ren Chen$^1$}

\affiliation{$^1$Neutron Scattering Division, Oak Ridge National Laboratory; Oak Ridge, TN 37831, USA}
\affiliation{$^2$Department of Nuclear Sciences and Engineering, Massachusetts Institute of Technology; Cambridge, MA 02139, USA}
\affiliation{$^3$Department of Engineering and System Science, National Tsing Hua University; Hsinchu 30013, Taiwan}
\affiliation{$^4$Institut Laue-Langevin; B.P. 156, F-38042 Grenoble Cedex 9, France}
\affiliation{$^5$Materials Science and Technology Division, Oak Ridge National Laboratory; Oak Ridge, TN 37831, USA}
\affiliation{$^6$Center for Nanophase Materials Sciences, Oak Ridge National Laboratory; Oak Ridge, TN 37831, USA}

\date{\today}

\begin{abstract}  

Small-angle neutron scattering (SANS) is a powerful technique for probing the nanoscale structure of materials. However, the fundamental limitations of neutron flux pose significant challenges for rapid, high-fidelity data acquisition required in many experiments. To circumvent this difficulty, we introduce a Bayesian statistical framework based on Gaussian process regression (GPR) to infer high-quality SANS intensity profiles from measurements with suboptimal signal-to-noise ratios (SNR). Unlike machine learning approaches that depend on extensive training datasets, the proposed \textit{one-shot} method leverages the intrinsic mathematical properties of the scattering function—smoothness and continuity—offering a generalizable solution beyond the constraints of data-intensive techniques. By examining existing SANS experimental data, we demonstrate that this approach can reduce measurement time by between one and two orders of magnitude while maintaining accuracy and adaptability across different SANS instruments. By improving both efficiency and reliability, this method extends the capabilities of SANS, enabling broader applications in time-sensitive and low-flux experimental conditions.

\end{abstract}

\maketitle

\clearpage


\section{Introduction}
\label{sec:introduction}

Small-angle neutron scattering (SANS) is an indispensable technique for investigating the structure of materials at the nanoscale. Neutrons offer unique advantages over other scattering probes, such as x-rays, including exceptional penetration power, negligible radiation damage, and the ability to selectively label specific material components through isotopic substitution \cite{ILL, JCNS}. These attributes have enabled significant advances in materials science, polymer science, and biology. However, unlike the remarkable progress in x-ray brightness achieved with synchrotron and free-electron laser sources, neutron brightness has remained largely stagnant in recent decades \cite{Narayanan}.  

This stagnation poses significant challenges for SANS experiments requiring rapid, high-fidelity data acquisition. Even the most advanced reactor- and spallation-based neutron sources, such as HFIR, ILL, SNS, and J-PARC, are fundamentally limited in their ability to increase neutron flux by orders of magnitude \cite{Granroth}. Consequently, obtaining high-quality SANS data often demands prolonged measurement times, ranging from tens of minutes to hours, even with state-of-the-art instrumentation. This limitation severely impedes real-time and \textit{in situ} investigations, such as studies on the structural evolution of mechanically driven materials. Moreover, low neutron flux restricts the exploration of weakly scattering systems and transient phenomena, where the relevant timescales are considerably shorter than the measurement duration imposed by neutron flux and the scattering power of the probed materials. These constraints highlight the critical need for innovative solutions \cite{Bruetzel, Lund}.  

Traditional approaches to addressing the flux bottleneck have focused on hardware upgrades, such as larger accelerators and brighter sources, but these solutions are often prohibitively expensive and logistically complex \cite{Taylor}. More recently, computational techniques, particularly machine learning (ML), have emerged as a promising alternative for accelerating data acquisition and analysis without increasing neutron flux \cite{Tennant}. For instance, deep learning-based super-resolution algorithms have demonstrated the ability to reconstruct high-resolution scattering data from measurements with limited detector counts, enabling faster experimental decision-making \cite{Shi, MCC}. However, these methods often require extensive training datasets collected at specific instruments, limiting their generalizability to other SANS configurations with different optical properties and performance characteristics.  

A promising yet underexplored approach for extracting robust information from sparse SANS measurements leverages the inherent mathematical properties of the scattering function. Theoretically, the scattering intensity, \(I(Q)\), is the Fourier transform of the autocorrelation function of the excess scattering length density profile, \(\gamma(r)\) \cite{ILL, Yip_NuclearRadiation}. In general soft matter systems, \(\gamma(r)\) lacks periodic long-range order, resulting in a smooth and continuous \(I(Q)\). Consequently, the intensity at any given \(Q\)-point, \(I(Q_i)\), inherently encodes information about its neighboring values, \(I(Q_{i-1})\) and \(I(Q_{i+1})\).  

Building on this property, we propose a Bayesian statistical approach \cite{jeffreys1939theory, gelman2013bayesian} to infer reliable SANS data from measurements with suboptimal signal-to-noise ratios (SNR). Bayesian inference is a probabilistic method that updates prior beliefs based on new evidence using Bayes' theorem \cite{bayes1763essay, jeffreys1939theory}. This approach has been widely applied across diverse fields, including predictive modeling in statistics \cite{gelman2013bayesian}, probabilistic learning in artificial intelligence \cite{bishop2006pattern}, clinical decision-making in medicine \cite{carlin2009bayesian}, ecological forecasting \cite{bolstad2007introduction, bauer2015quiet, carrassi2018data}, and inverse problems in scattering \cite{Bayesian1, Bayesian2, Bayesian3, GPRdiblock, GPRcolloid, GPRladder, GPRcharge, GPRmechanical}. However, existing Bayesian methods rely on priors derived from extensive databases or analytical models, limiting their applicability to well-documented systems. In the context of sparse SANS measurements, where prior knowledge of the probed material is often unavailable, pre-training on large datasets becomes impractical.  

Rather than relying on material- or instrument-specific priors, we introduce a Bayesian approach that begins with a non-informative prior, reflecting the fundamental characteristics of \( I(Q) \): smoothness and continuity. This prior is formulated based on correlations between data points and refined through Gaussian process regression (GPR) \cite{williams1996gaussian, williams1998prediction, GPR, Deringer2021Gaussian, Noack2021Gaussian}, which incorporates experimental measurements, \( I_{\mathrm{Exp}}(Q) \), to infer the posterior distribution. The resulting posterior preserves smoothness and continuity while effectively suppressing random noise in \( I_{\mathrm{Exp}}(Q) \), enabling the reconstruction of high-fidelity SANS data even from measurements with insufficient neutron flux. Computational benchmarking and experimental SANS data validation demonstrate that this method can enhance data quality and reduce measurement times by one to two orders of magnitude, while maintaining adaptability across different SANS instrument configurations.

Although increasing neutron source flux or improving optical components could, in principle, enhance data collection, such hardware-based solutions are often impractical due to their high costs, technical complexity, and long development cycles. In contrast, the proposed statistical inference framework provides a cost-effective and scalable alternative, significantly improving data quality without requiring modifications to the experimental setup.  

This adaptive Bayesian inference framework overcomes the limitations of traditional approaches that rely on system-dependent priors, making it broadly applicable to a wide range of materials, including those with limited prior characterization. By enabling robust inference across diverse systems, it enhances neutron scattering methodologies for high-throughput and time-sensitive studies. This advancement significantly expands the scope of SANS, providing a more versatile and efficient approach to probing nanoscale structures across various scientific disciplines.  

The remainder of this paper is organized as follows. In Section~\ref{sec:methods}, we present the methods used to develop the GPR-based statistical inference for obtaining robust scattering intensity from sparse measurements. In Section~\ref{sec:results}, we examine the feasibility of the proposed approach for experimental SANS measurements. Finally, in Section~\ref{sec:conclusion}, we discuss the results and conclude with a summary of the work and potential future directions.  

\section{Methods}
\label{sec:methods}

\subsection{Visualizing Statistical Precision in SANS Measurements}

Before delving into the mathematical details of statistical inference for SANS measurements, we first present an illustrative representation of the problem at hand. Such a visualization provides intuitive insight into the key challenges and objectives, offering a conceptual foundation that aids in understanding the subsequent analytical developments.

\begin{figure}[h!]
\centerline{
  \includegraphics[width =\columnwidth]{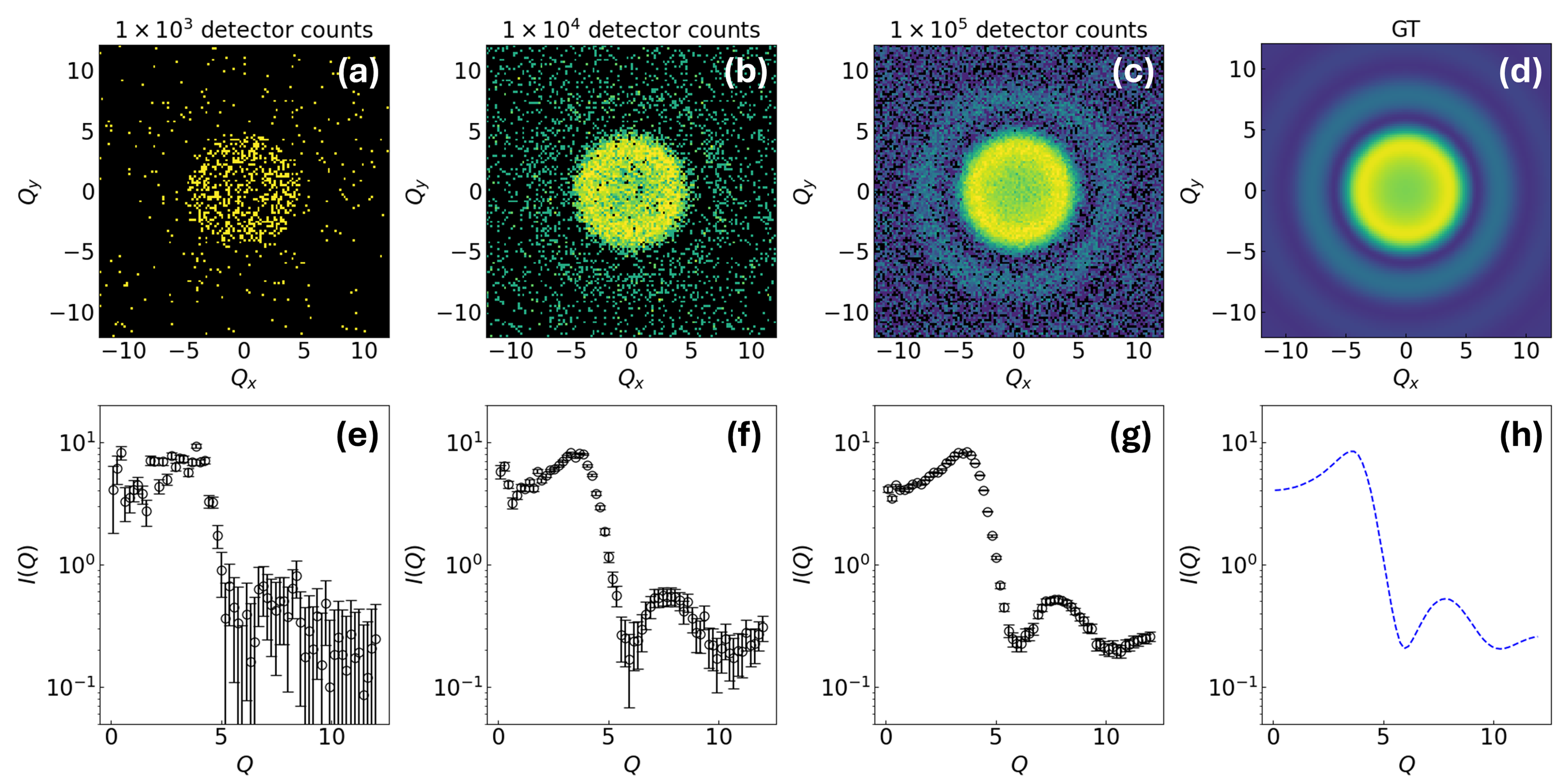}
}  
\caption{Impact of increasing detector counts on two-dimensional computationally synthetic spectra and radially averaged intensity profiles. (a)–(c) show two-dimensional scattering patterns collected with $3 \times 10^3$, $1 \times 10^4$, and $3 \times 10^4$ neutrons, respectively, while (d) represents the ground truth (GT) pattern. The corresponding radially averaged intensity curves, shown in (e)–(g), exhibit reduced statistical noise with increasing detector counts. (h) depicts the reference intensity curve derived from the GT data.}
\label{fig:1}
\end{figure}

Fig.~\ref{fig:1} illustrates the impact of increasing detector counts on the quality of two-dimensional computationally synthetic scattering spectra and their corresponding radially averaged intensity profiles. Panels (a)–(c) show scattering patterns obtained with detector counts of \(1 \times 10^3\), \(1 \times 10^4\), and \(1 \times 10^5\), respectively. In contrast, panel (d) displays the intensity distribution computed directly from the theoretical scattering function of interacting hard spheres at a volume fraction of 0.3. The scattering intensity in this system is represented as the product of the hard sphere form factor, \( P(Q) \), and the inter-particle structure factor, \( S(Q) \), derived by solving the Ornstein-Zernike Eqn.~\cite{OZ} with the Percus-Yevick closure~\cite{PY}. This computation follows Baxter's analytical approach~\cite{Baxter1, Baxter2, Hansen}, further refined by the Wertheim correction~\cite{Wertheimt} to account for structural collectivity. The theoretically computed scattering spectrum in panel (d) serves as the ground truth (GT).

As the detector counts increase, the SNR improves substantially, providing a more accurate representation of the underlying scattering patterns. This enhancement reduces statistical noise, revealing finer structural details with greater clarity. Consequently, previously obscured features in the scattering spectra become more discernible, enabling more precise quantitative regression analysis and a deeper interpretation of the system's structural properties.

This trend is evident in the radially averaged one-dimensional representations shown in panels (e)–(h). As indicated in panel (h), the theoretically calculated, noise-free \( I(Q) \) is smooth and continuous. For clarity, in this report, \( I_\mathrm{GT}(Q) \) is used to denote the ideal, noise-free scattering intensity. However, limited instrument flux or very short measurement times can result in noisy data. Simulations of neutron arrivals at the detector, with counts ranging from \(10^3\) to \(10^5\), illustrate this behavior.

At low detector counts, the scattering intensity exhibits considerable uncertainty following Poisson statistics \cite{kingman1993poisson}, where the relative uncertainty in intensity is described by:
\begin{equation}
    \frac{\Delta I}{I} \propto \frac{1}{\sqrt{N}},
\end{equation}
with \( N \) representing the total neutron count. As the neutron count increases from \(10^3\) (panel (e)) to \(10^5\) (panel (g)), the uncertainty decreases, resulting in smoother intensity profiles that more closely match the reference curve in panel (h). This inverse square-root dependence means that an orders-of-magnitude increase in neutron count is required to yield more accurate intensity profiles and greater reliability in the structural information extracted from scattering data. However, as noted in Section~\ref{sec:introduction}, the increase in neutron flux is hard to achieve. This fundamental limitation emphasizes the importance of establishing a reliable inference from low-count data.  

\subsection{Leveraging Gaussian Process Regression GPR for Statistical Inference}

\begin{figure}[h!]
\centerline{
  \includegraphics[width=0.75\textwidth]{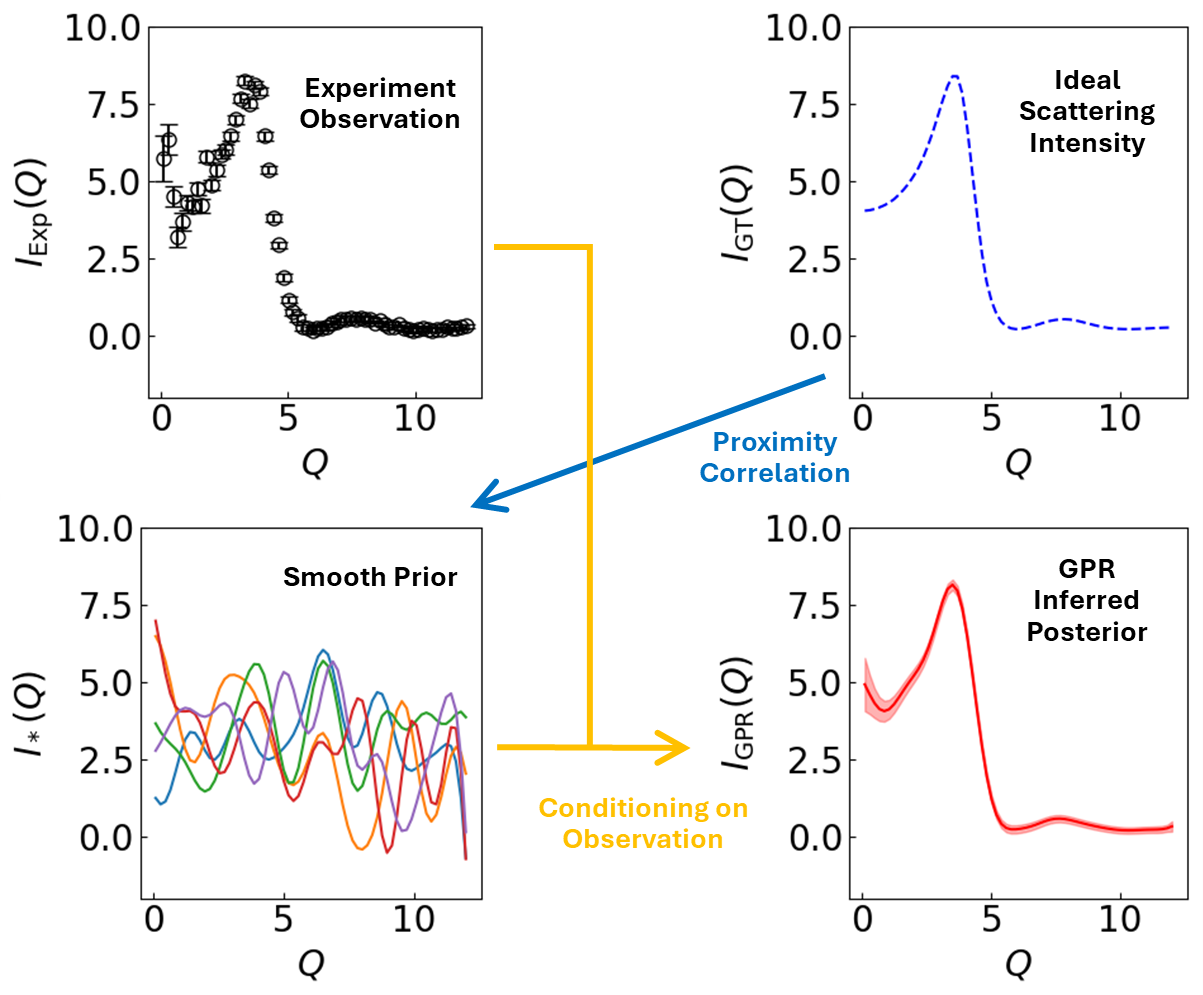}
}  
\caption{Probabilistic framework for reconstructing smooth and continuous scattering intensity \( I_\mathrm{GT}(Q) \) using Gaussian process regression (GPR). The top-left panel shows the experimental scattering intensity \( I_\mathrm{Exp}(Q) \) affected by measurement noise. The bottom-left panel illustrates prior realizations of \( I(Q) \) drawn from a multivariate normal distribution, enforcing smoothness through a covariance structure. The top-right panel presents the ground truth noise-free scattering intensity \( I_\mathrm{GT}(Q) \). The bottom-right panel displays the reconstructed intensity \( I_*(Q) \) with uncertainty estimates, demonstrating the effectiveness of the GPR framework in suppressing noise while preserving signal fidelity.}
\label{fig:2}
\end{figure}

Our objective is to construct a probabilistic framework that preserves the smoothness and continuity of \( I_\mathrm{GT}(Q) \) while incorporating experimental data (Fig.~\ref{fig:2}). First, the scattering intensity is represented by a multivariate prior distribution that enforces these properties, offering a flexible model before data incorporation.

To formalize this prior, we assume the ground truth pattern fluctuates around some smooth, \(Q\)-dependent reference value $m(Q)$, which can be approximated by applying either Gaussian smoothing or Savitzky–Golay filtering \cite{press2007numerical} to the measured $I_\mathrm{Exp}$.

Furthermore, we model the correlation between scattering intensities \( I_1-m_1 \) and \( I_2-m_2 \) at different \( Q \)-values using a bivariate normal distribution, with the correlation structure captured by a covariance matrix:
\begin{equation}
K = \begin{bmatrix}
k_{11} & k_{12} \\
k_{21} & k_{22}
\end{bmatrix},
\label{eq:2}
\end{equation}
where $m_i=m(Q_i)$, \( k_{11} \) and \( k_{22} \) represent the variances of \( I_1-m_1 \) and \( I_2-m_2 \), and \( k_{12} \) and \( k_{21} \) capture their correlation. 

This framework extends to an \( N \)-dimensional multivariate normal distribution with a correlation matrix:
\begin{equation}
K = \begin{bmatrix}
k_{11} & \cdots & k_{1N} \\
\vdots & \ddots & \vdots \\
k_{N1} & \cdots & k_{NN}
\end{bmatrix}.
\label{eq:3}
\end{equation}
Each entry \( k_{ij} \) reflects the statistical dependence between \( I(Q_i) \) and \( I(Q_j) \). To ensure smoothness while allowing correlation to decay over distance, we define the covariance as:
\begin{equation}
k_{ij} = k(Q_i, Q_j) = \alpha\exp\left(-\frac{\|Q_i - Q_j\|^2}{2\lambda^2}\right),
\label{eq:4}
\end{equation}
where \( \lambda \) governs the correlation length, ensuring strong correlation for nearby intensities persists and decays progressively as the distance between \( Q_i \) and \( Q_j \) increases. Note that the selection of an optimized \( \lambda \) involves complex factors such as the functional form of \( I(Q) \) and the density of \( Q \) points. In this work, we take an uninformative approach on purpose and choose not to explore the specifics of optimizing \( \lambda \). Since the unit of \( k_{ij} \) should be the squared unit of \( I \), we set the prefactor of the exponential function, \( \alpha \), to be the signal variance, defined as
\begin{equation}
\sigma_f^2 = \frac{\sum_{i \in \{1:n_Q\}} (I_i-m_i)^2}{n_Q}.
\label{eq:5}
\end{equation}

According to \(m\) and \(K\), the prior distribution of a scattering function \( I \) is:
\begin{equation}
p(I) = \mathcal{N}(m,K),
\label{eq:6}
\end{equation}
where \( \mathcal{N} \) denotes the normal distribution. Eqn.~\eqref{eq:6} indicates that the prior distribution \( P(I) \) fluctuates around a given reference \( m \), with smoothness enforced by the covariance matrix \( K \) following Eqn.~\eqref{eq:5}. The choice of \( m \) plays a crucial role in data interpretation and will be discussed further. Therefore, \( \sigma_f^2 \) can be regarded as a measure of the magnitude of random noise, closely approximating the average experimental uncertainty.

The fact that the measured data, denoted as \( I_{\mathrm{Exp}} \), and the smooth output, denoted as \( I_* \) evaluated at \( Q_* \), follow a joint multivariate normal distribution enables us to formulate a mathematical expression to incorporate experimental observation \( I_\mathrm{Exp}\) in the context of GPR~\cite{rasmussen2006gaussian, bishop2006pattern, murphy2012machine, gelman2013bayesian}:

\begin{equation}
p\begin{pmatrix}
I_\mathrm{Exp} \\
I_*
\end{pmatrix} = 
\mathcal{N}\left(\begin{bmatrix}
m \\
m_*
\end{bmatrix},
\begin{bmatrix}
K & K_{*}^\top \\
K_{*} & K_{**}
\end{bmatrix}\right).
\label{eq:7}
\end{equation}
According to Eqn.~\eqref{eq:5}, the joint distribution covariance matrices \( K_{*} \) and \( K_{**} \) can be constructed in terms of \( K \) by iterating over pairs \( (Q_i, Q_j) \), where \( K_{*} \) corresponds to \( (Q, Q_*) \) and \( K_{**} \) corresponds to \( (Q_*, Q_*) \). The value of \( m_* \) can be obtained by interpolating \( m \), leveraging its inherent smoothness.

With the availability of experimental observation \( I_{\mathrm{Exp}} \), we can update our belief of \( I_* \) as \( I_\mathrm{GPR} \) based on the following conditional distribution:

\begin{equation}
I_\mathrm{GPR} \sim p(I_* | I_\mathrm{Exp}, Q, Q_*) = \mathcal{N}(\mu_*, \sigma^2_*),
\label{eq:8}
\end{equation}
where
\begin{equation}
\mu_* = K_{*}K^{-1}(I_\mathrm{Exp}-m)+m_*
\label{eq:9}
\end{equation}
represents the mean of the post-experiment conditional distribution, and
\begin{equation}
\sigma^2_* = K_{**}-K_{*}K^{-1}K_{*}^\top
\label{eq:10}
\end{equation}
is the corresponding covariance matrix. The appropriate standard deviation for \( I_* \) can be assigned as

\begin{equation}
\Delta I_\mathrm{GPR}(Q_i) = \sqrt{\sigma^2_{*,ii}}.
\label{eq:11}
\end{equation}

The process of implementing the GPR-based statistical inference for SANS sparse measurements is systematically outlined in Fig.~\ref{fig:2}. This framework leverages GPR to enhance measurement accuracy, providing a robust approach for analyzing sparse SANS datasets.

\subsection{Uncertainty Estimations from Noisy Input: Experimental Data Connection}

The mathematical formulation of Eqns.~\eqref{eq:8}-\eqref{eq:10} assumes that each experimental observation is noise-free. However, as illustrated in the top-left panel of Fig.~\ref{fig:2}, experimental measurements inherently include uncertainties due to statistical fluctuations, detector limitations, and other sources of noise. This implies that different data points carry varying levels of confidence in influencing modifications to the prior. To account for these uncertainties, we assume that the noise terms are independently and identically distributed (i.i.d.) and proportional to the magnitude of experimental uncertainties $\Delta I(Q)$. Based on this assumption, the covariance matrix of the experimental input in Eqn.~\eqref{eq:7} can be reformulated as

\begin{equation}
K' = K + \Sigma,
\label{eq:12}
\end{equation}
an additional diagonal term $\Sigma$ has been inserted
\begin{equation}
\Sigma = \mathrm{diag}\left(m_\sigma^2[\Delta I_\mathrm{Exp}(Q)]^2\right).
\label{eq:13}
\end{equation}
The predicted mean and variance is given by
\begin{equation}
\mu_* = K_*(K+\Sigma)^{-1}(I_\mathrm{Exp}-m)+m_*,
\label{eq:16}
\end{equation}
\begin{equation}
\sigma^2_* = K_{**}-K_{*}(K+\Sigma)^{-1}K_{*}^\top.
\label{eq:16a}
\end{equation}
Here, \( m_\sigma \) is a dimensionless scaling factor that reflects the relationship between independent noise and the SNR of the experimental data. Since the signal variance, as defined in Eqn.~\ref{eq:5}, is influenced by the smooth background or prior mean, it is not possible to determine \( m_\sigma \) from \( I_\mathrm{Exp} \) a priori.  
To address this, we propose an evidence-driven calibration method to determine the most appropriate value of \( m_\sigma \):
\begin{equation}
m_\sigma = \underset{m_\sigma}{\operatorname{argmax}} \ L(m_\sigma; I_\mathrm{Exp}, p(I_* | I_\mathrm{Exp}, Q, Q_*)).
\label{eq:14}
\end{equation}
Here, $L$ represents the log-likelihood of the observed data given the posterior distribution from GPR, expressed as:
\begin{equation}
L = \frac{1}{n_Q}\sum_{i=1}^{n_Q} \left[ -\frac{1}{2} \log(2\pi\Delta I_*^2) - \frac{(I_\mathrm{Exp} - \mu_*)^2}{2\Delta I_*^2} \right].
\label{eq:15}
\end{equation}

\begin{figure}[!h]
    \centering
    \includegraphics[width=1.0\columnwidth]{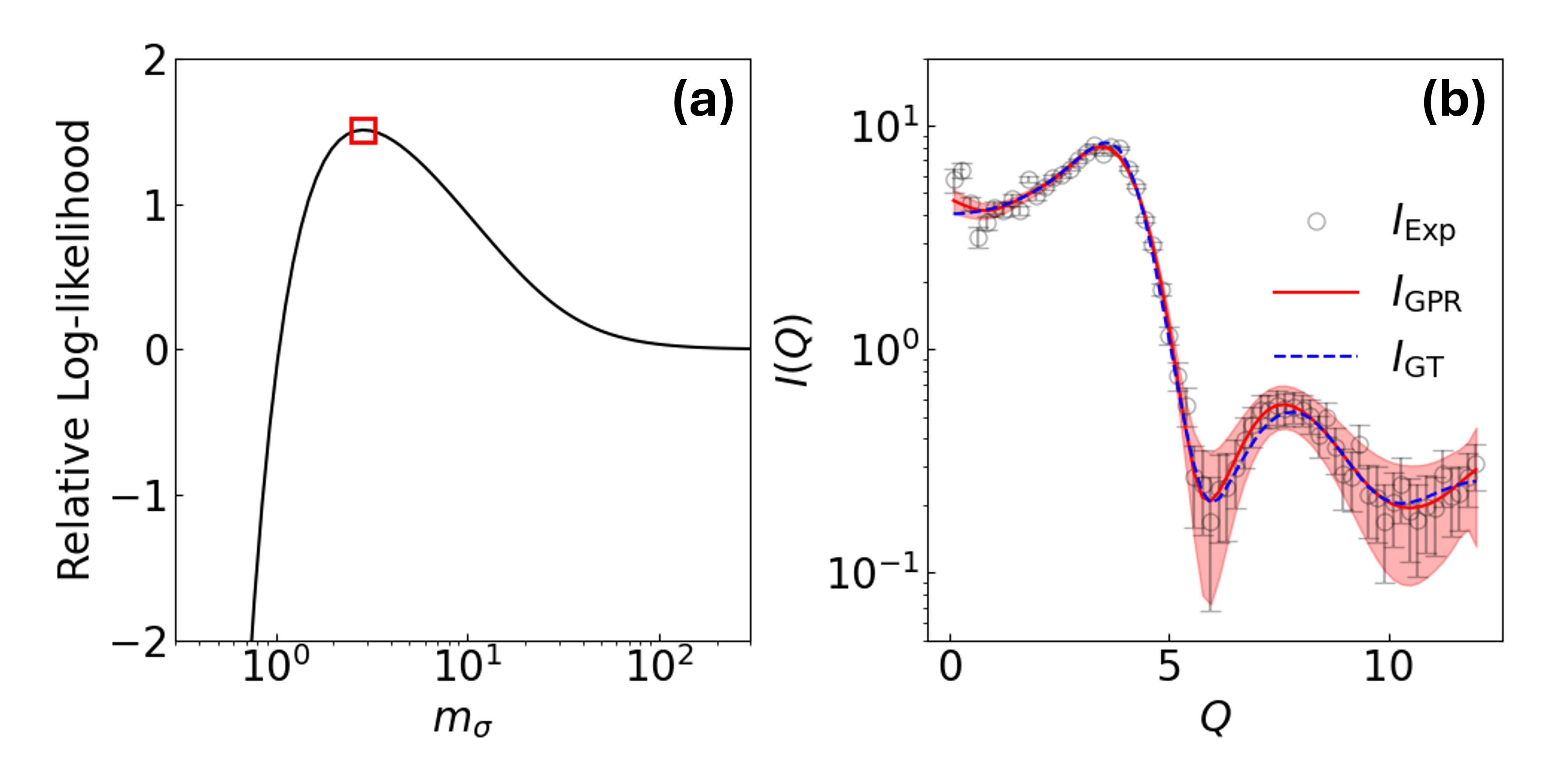}
    \caption{(a) Behavior of $L$ for the hard-sphere system at a volume fraction of 0.3 as a function of $m_\sigma$. The selected scattering intensity for illustration corresponds to Fig.~\ref{fig:2}(f), with detector counts of $1 \times 10^4$. The optimized value of $m_\sigma$ is indicated by the red square. (b) Posterior distribution, derived from Eqns.~\eqref{eq:9} and \eqref{eq:12} using the optimized value of $m_\sigma$, represented by the red shaded region. Strong quantitative agreement is observed between the inferred mean (red curve) and $I_{\text{GT}}$.}
    \label{fig:3}
\end{figure}
Fig.~\ref{fig:3}(a) depicts the behavior of $L$ for the hard-sphere system at a volume fraction of 0.3 as a function of $m_\sigma$. The selected scattering intensity for illustration corresponds to Fig.~\ref{fig:2}(f), with detector counts of $1 \times 10^4$. The posterior distribution, derived from Eqns.~\eqref{eq:9} and \eqref{eq:12} using the optimized value of $m_\sigma$, is represented by the red shaded region in Fig.~\ref{fig:3}(b), while the optimized value itself is indicated by the red square in Fig.~\ref{fig:3}(a). A strong quantitative agreement is observed between the inferred mean (red curve) and the reference intensity $I_{\text{GT}}$.
\begin{figure}[h]
    \centering
    \includegraphics[width=0.5\columnwidth]{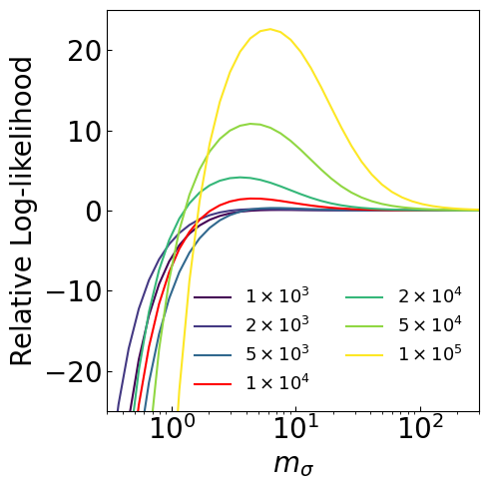}
    \caption{Log-likelihood as a function of $m_{\sigma}$ for different total detector counts, illustrating the impact of noise on the dataset. The variations in the log-likelihood highlight how the level of noise influences the accuracy and precision of the inferred parameters.}
    \label{fig:4}
\end{figure}

Before proceeding further, it is instructive to examine the behavior of the likelihood as a function of $m_{\sigma}$ at different detector counts. As shown in Fig.~\ref{fig:4}, the log-likelihood increases monotonically for cases with relatively low total detector counts (1000–3000) without exhibiting a maximum. The noise component $\Sigma$ dominates the observed data distribution in such scenarios. Given that the prediction and observation share identical $Q$-points ($Q_* = Q$), an eigendecomposition of $K_*=K$ in Eqn.~\eqref{eq:16} provides:
\begin{equation}
K = \sum_{i=1}^{n_Q} \Lambda_i^2\mathbf{u_i}\mathbf{u_i}^\top,
\label{eq:17}
\end{equation}
where $\Lambda_i$ and $\mathbf{u}_i$ represent the eigenvalues and eigenvectors, respectively. The observation $(I_\mathrm{Exp} - m)$ can be projected onto $\mathbf{u}_i$ as:
\begin{equation}
I_\mathrm{Exp} - m = \sum_{i=1}^{n_Q} \gamma_i \mathbf{u}_i.
\label{eq:18}
\end{equation}
Substituting Eqns.~\eqref{eq:17} and \eqref{eq:18} into Eqn.~\eqref{eq:16}, we find:
\begin{equation}
\mathbf{\mu} = m + \sum_{i=1}^{n_Q} \frac{\gamma_i\Lambda_i^2}{\Lambda_i^2 + \Sigma_i^2} \mathbf{u_i},
\label{eq:19}
\end{equation}
where $\Sigma_i^2 = m_\sigma^2(\Delta I(Q_i))^2$. When $\frac{\Lambda_i^2}{\Lambda_i^2 + \Sigma_i^2} \ll 1$, the contribution of $I_\mathrm{Exp}$ becomes negligible, reducing $\mu$ to $m$ as per Eqn.~\eqref{eq:19}. In this case, the observed data provides no additional information beyond the prior distribution, meaning that inference remains heavily dependent on prior assumptions. The presence of a notable maximum in the log-likelihood at sufficiently low $m_\sigma$ thus serves as an indicator of whether the collected neutron count is sufficient to yield meaningful information.

In addition to incorporating the uncertainty in the measured scattering intensity through Eqn.~\eqref{eq:12}, it is also important to consider the effect of instrument resolution \cite{Pedersen2025}. A discussion on how a given resolution function influences the covariance matrix is provided in Appendix~\ref{app:a}.

\subsection{Computational Benchmarking: Synthetic Data
}

Before assessing the approach with experimentally obtained SANS data, it is crucial to establish a computational benchmark to evaluate the effectiveness of statistical inference. Since SANS serves as a key technique for probing the structure of soft materials, we focus on three major categories that are widely recognized in soft matter research from a scattering perspective \cite{ILL} to ensure a robust validation framework.

\begin{figure}[h]
    \centerline{
    \includegraphics[width = 1.0\columnwidth]{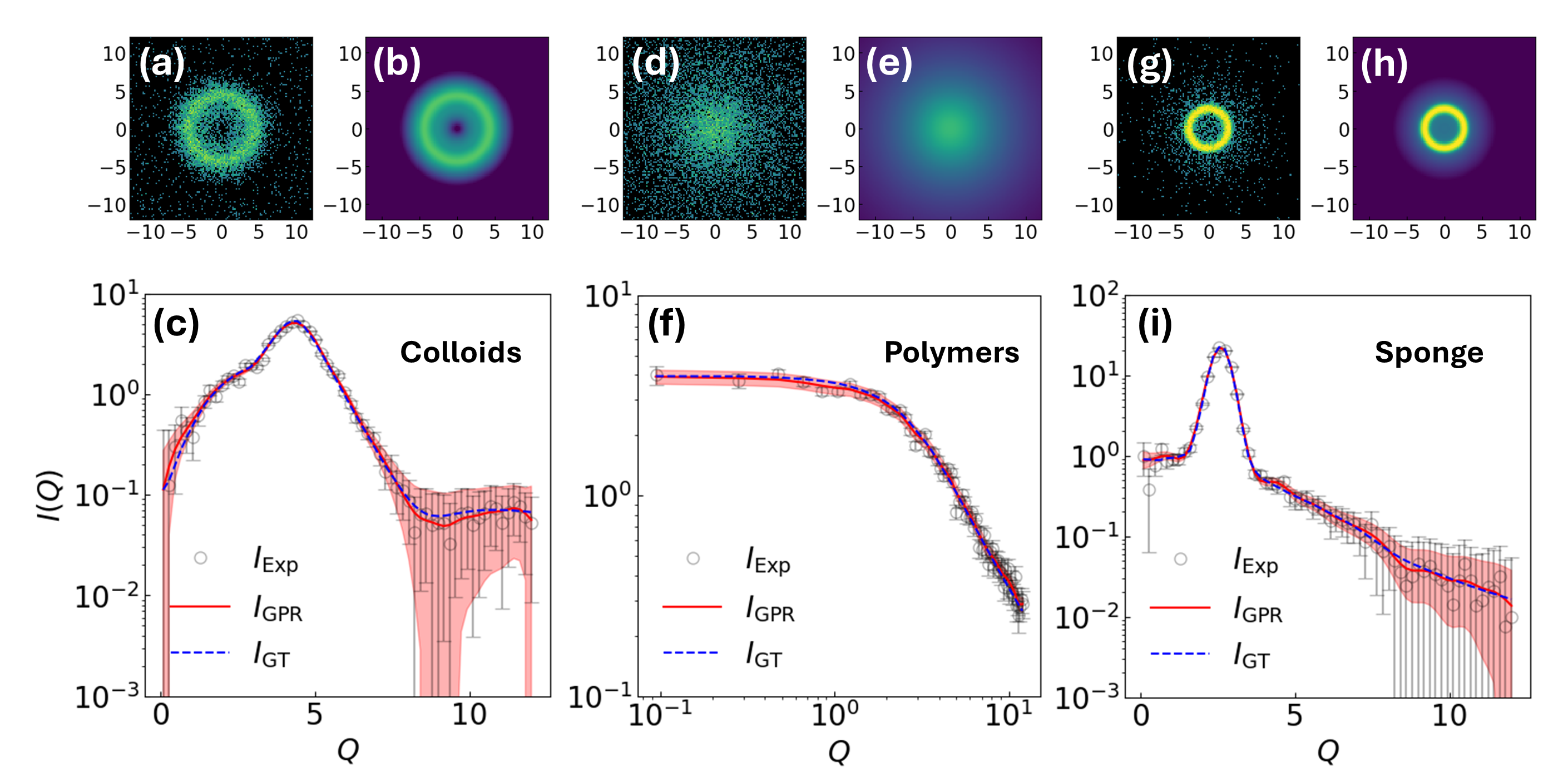}
    }
\caption{  
(a, b) Two-dimensional computationally synthesized scattering spectra of a charge-stabilized colloidal suspension (volume fraction 0.3), calculated with total detector counts of $10^{4}$ and infinite counts, respectively.   
(c) Azimuthally averaged one-dimensional intensity profiles: (dashed line) theoretical ground truth intensity $I_\mathrm{GT}$, (circles) noisy experimental intensity $I_\mathrm{Exp}$ from (a), (shaded region) GPR-predicted confidence interval $I_\mathrm{GPR}$, and (solid line) mean of $I_\mathrm{GPR}$.  
(d, e) Scattering spectra of a continuous Kratky-Porod (KP) chain~\cite{KP1, KP2, Landau} under the same detector count conditions as (a, b).   
(f) Comparison of $I_\mathrm{GPR}$ confidence intervals with uncertainties in $I_\mathrm{Exp}$, demonstrating agreement between the mean of $I_\mathrm{GPR}$ and $I_\mathrm{GT}$ over the probed $Q$ range.   
(g, h) Scattering spectra of an isotropic sponge phase, computed using Berk’s leveled wave field model~\cite{Berk1, Berk2}.   
(i) Validation of GPR-based statistical inference for lyotropic phases.  
}

    \label{fig:5}
\end{figure}

The first representative class we consider is colloidal suspensions composed of globular particles. Many soft materials, including various colloids, self-assembled micellar systems, and protein solutions, can be broadly categorized within this group in the context of scattering data analysis. In interacting systems, different forms of coarse-grained interactions arise \cite{Dhont}, such as hard-sphere interactions, screened Coulombic repulsion, and steric repulsion, all of which are commonly observed in experimental systems. A theoretical isostructurality condition has been established \cite{Andersen, Verlet, Hansen}, demonstrating that when particles interact through a purely repulsive, centrosymmetric potential, their static structure—characterized by the two-point static correlation, the primary quantity of interest in SANS—can be mapped onto that of a hard-sphere system with an effective hard-sphere diameter and an effective volume fraction. Based on this, we select interacting hard-sphere solutions as the first system to evaluate the efficacy of statistical inference, as their structural characteristics are representative of many soft materials within this classification.

Fig.~\ref{fig:5}(a) displays the two-dimensional scattering spectrum of a charge-stabilized colloidal suspension at a volume fraction of 0.3, obtained with a total detector count of $10^{4}$. In contrast, Fig.~\ref{fig:5}(b) presents the ideal two-dimensional scattering spectrum computed assuming an infinite detector count. The radially averaged one-dimensional scattering intensity in Fig.~\ref{fig:5}(c) compares the theoretical intensity ($I_\mathrm{GT}$, blue dashed curve) as the ground truth. The gray symbols ($I_\mathrm{Exp}$) represent the intensity derived from the finite-count spectrum, exhibiting significant uncertainties. The shaded region denotes the confidence interval of the intensity predicted by GPR ($I_\mathrm{GPR}$), with the red curve representing the mean prediction, which closely aligns with $I_\mathrm{GT}$.

Figs.~\ref{fig:5}(d) and (e) display the two-dimensional scattering spectra of a continuous Kratky-Porod (KP) chain~\cite{KP1, KP2, Landau}, a reference model in computational studies of semiflexible polymers~\cite{Binder1, Binder2}. The spectra are calculated for a total detector count of $10^{4}$ and an infinite number of detector counts. The KP chain is chosen for its relevance in describing the scattering characteristics of various polymer systems, including dilute or interacting semiflexible self-avoiding linear chains and branched structures~\cite{ILL}. Fig.~\ref{fig:5}(f) compares these spectra, showing that, similar to Fig.~\ref{fig:5}(c), the confidence interval of $I_\mathrm{GPR}$ is comparable to the uncertainties of $I_\mathrm{Exp}$, and the mean of $I_\mathrm{GPR}$ exhibits quantitative agreement with $I_\mathrm{GT}$ over the probed $Q$ range.

Figs.~\ref{fig:5}(g) and (h) present the two-dimensional scattering spectra of an isotropic sponge phase, calculated using the leveled wave field model proposed by Berk~\cite{Berk1, Berk2}, under the same two detector count conditions. Other lyotropic systems, such as distorted lamellar phases with varying topological defects and multilayer onion structures, can also be described within this framework~\cite{Tung}. The applicability of GPR-based statistical inference for general lyotropic phases is further supported by the results shown in Fig.~\ref{fig:5}(i).

The examples above demonstrate that SANS spectra of soft matter exhibit highly diverse behavior, making ML-based image processing methods that rely on external training sets highly unsuitable for quantitive study of these systems. Most image processing techniques are optimized for human perception, which has a limited dynamic range (typically below three orders of magnitude)~\cite{Darmont_humanvision}, and is better suited for detecting uniform patterns or gradual variations~\cite{Klein_humanvision}. In contrast, SANS spectra can feature extreme intensity variations, with the ratio between peak and valley scattering intensities often exceeding three to four orders of magnitude. Meanwhile, the spectra of polymer chains exhibit more gradual variations around the inverse of their radius of gyration. Regarding characteristic length scales, the peak width of lyotropic systems can be as narrow as one-tenth of the peak position, while the rest of the spectrum remains nearly flat. The fact that these highly system-dependent structural features cannot be effectively captured by a predefined, universal training set thereby underscores the advantages of the proposed one-shot inference method.

\subsection{Quantitative Analysis: Efficiency Improvements through GPR}

As an illustrative example, the scattering intensities of a hard sphere system with a volume fraction of 0.3, presented in Fig.~\ref{fig:1}, are used to evaluate the efficiency of GPR-based statistical inference. To assess the accuracy of the GPR predictions in capturing the ground truth distribution, we define the target distribution around $I_\mathrm{GT}$ as
\begin{equation}
    p_{\mathrm{GT}} = \mathcal{N}(I_{\mathrm{GT}},\sigma_f^2).
 \label{eq:21}
\end{equation}
Next, we compute the relative entropy, also known as the Kullback–Leibler divergence~\cite{KL}, between any given distribution $p$ and the ground truth distribution $p_{\mathrm{GT}}$:
\begin{equation}
   D(p_\mathrm{GT}||p) = H(p_\mathrm{GT}, p)-H(p_\mathrm{GT})
   \label{eq:22}
\end{equation}
where $H$ denotes the cross entropy
\begin{equation}
    H(p_\mathrm{GT}, p) = -\int dI \ p_\mathrm{GT}(I) \log p(I),
\end{equation}
\begin{equation}
    H(p_\mathrm{GT})\equiv H(p_\mathrm{GT}, p_\mathrm{GT}) = -\int dI \ p_\mathrm{GT}(I) \log p_\mathrm{GT}(I).
\end{equation}
\( D(p_\mathrm{GT}||p) \), measured in nats (with the logarithm taken to base \( e \)), quantifies the divergence between the predicted distribution \( p \) and the target distribution \( p_{\mathrm{GT}} \). It indicates the amount of information lost when \( p \) is used as an approximation of \( p_{\mathrm{GT}} \). Given a random variable following \( p_{\mathrm{GT}} \), its likelihood of belonging to \( p \) is reduced relative to its likelihood of belonging to \( p_{\mathrm{GT}} \) by a factor of \( \exp(-D(p_\mathrm{GT}||p)) \). A lower cross-entropy value signifies a closer match between the predicted and actual distributions, whereas higher values indicate greater divergence. As \( p \) approaches \( p_{\mathrm{GT}} \), the relative entropy tends to zero, i.e., \( D(p_\mathrm{GT}||p) \to 0 \). To quantify prediction accuracy, we set the variance of \( p_{\mathrm{GT}} \) to the signal variance in Eqn.~\ref{eq:21}. Under this choice, when \( D(p_\mathrm{GT}||p) = 1 \), the mismatch of the given prediction \( p \) from \( p_{\mathrm{GT}} \) is on the same order as the deviation between \( I_{\mathrm{GT}} \) and \( m \). This deviation is expected to be significantly smaller than the uncertainty caused by noise resulting from insufficient statistical sampling, provided that \( m \) is properly chosen to be encapsulated within the spread of \( I_\mathrm{Exp} \).

We compute $D(p_\mathrm{GT}||p)$ averaged over the experimental $Q$-range for the following distributions, including the GPR posterior as stated in Eqns.~\eqref{eq:16}-\eqref{eq:16a}
\begin{equation}
    p_{\mathrm{GPR}} = p(I_* | I_{\mathrm{Exp}}, Q, Q_*) = \mathcal{N}(\mu_*, \sigma^2_*),
    \label{eq:p_GPR}
\end{equation}
and the raw distribution from measurement
\begin{equation}
    p_{\mathrm{Exp}} = \mathcal{N}(I_{\mathrm{Exp}}, \Delta I_{\mathrm{Exp}}).
    \label{eq:p_Exp}
\end{equation}
For simplicity, the following \( Q \)-averaged divergences are defined based on the distributions given in Eqns.~\eqref{eq:p_GPR}-\eqref{eq:p_Exp}:  

\begin{equation}
    D_{\mathrm{GPR}} = \left\langle D(p_{\mathrm{GT}}||p_{\mathrm{GPR}})\right\rangle_{Q},
\end{equation}
\begin{equation}
    D_{\mathrm{Exp}} = \left\langle D(p_{\mathrm{GT}}||p_{\mathrm{Exp}})\right\rangle_{Q}.
\end{equation}
\begin{figure}[h]
    \centering
    \includegraphics[width=1.0\textwidth]{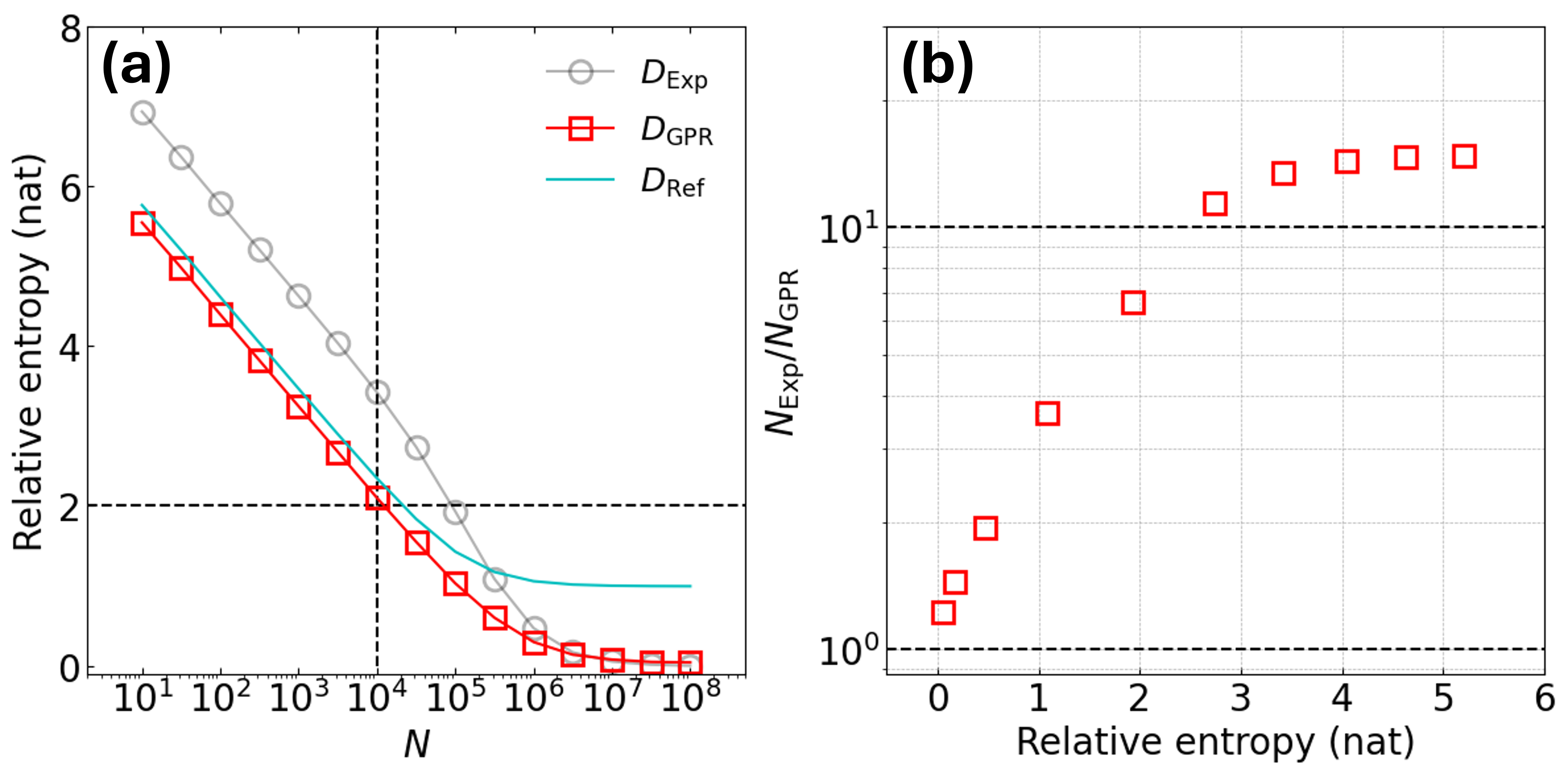}  
    \caption{(a) Relative entropy \( D(p_{\mathrm{GT}}||p) \) for three distributions: GPR-based posterior \( p_{\mathrm{GPR}} \), experimental distribution \( p_{\mathrm{Exp}} \), and reference prior \( p_{\mathrm{Ref}} \). As the measurement count increases, noise is suppressed, and both \( D(p_{\mathrm{GT}} \parallel p_{\mathrm{GPR}}) \) and \( D(p_{\mathrm{GT}} \parallel p_{\mathrm{Exp}}) \) approach zero, indicating improved agreement with the ground truth. \( D_{\mathrm{Ref}} \) exhibits a plateau for \( N>10^6 \), reflecting distortions in the spectrum caused by the smoothing procedure. The vertical dashed line highlights a 2-nat accuracy improvement, from \( p_{\mathrm{Exp}} \) (4 nats) to \( p_{\mathrm{GPR}} \) (2 nats), at a detector count of \( N = 10^4 \), as exemplified by the interacting hard sphere system in Fig.~\ref{fig:3}. Achieving this level of improvement without GPR would require an order of magnitude more neutrons, i.e., \( N = 10^5 \). For visual reference, readers can consult the synthetic spectra shown in Figs.~\ref{fig:1} and \ref{fig:3}.
    }
    \label{fig:6}
\end{figure}
In Fig.~\ref{fig:6}(a), for \(N<3\times10^5\), \( D_{\mathrm{Exp}}>1\), implying the signal is significantly distorted by noise. The relative entropy for all three distributions decreases as the neutron count increases. This reduction can be attributed to noise suppression, in accordance with the central limit theorem. As the system approaches the long-time limit, both \( D_{\mathrm{Exp}} \) and \( D_{\mathrm{GPR}} \) converge to zero, indicating that the predictions from \( p_{\mathrm{GPR}} \) and \( p_{\mathrm{Exp}} \) become indistinguishable within the range of \( p_{\mathrm{GT}} \).

To illustrate this further, consider the example shown in Fig.~\ref{fig:3}. For the experimental data with \( N = 10^4 \), the relative entropy \( D_{\mathrm{Exp}} \) is approximately 4 nats, which corresponds to the noisy case presented in Figs.~\ref{fig:1}(b) and (e). As highlighted by the vertical dashed line in Fig.~\ref{fig:6}(a), the application of GPR reduces the relative entropy by roughly 2 nats, thereby achieving the same data quality as the raw experimental distribution in Figs.~\ref{fig:1}(c) and (f). This level of data quality would otherwise require an order of magnitude more neutron flux to obtain, as inferred from the horizontal dashed line in Fig.~\ref{fig:6}(a).

The cyan solid line in Fig.~\ref{fig:6}(a) represents \( D_{\mathrm{Ref}} = \left\langle D(p_{\mathrm{GT}}||p_{\mathrm{Ref}}) \right\rangle_{Q} \), where \( p_{\mathrm{Ref}} \) denotes the prior distribution as defined in Eqn.~\ref{eq:6}. Notably, the gap between \( D_{\mathrm{Ref}} \) and \( D_{\mathrm{Exp}} \) for \( N < 10^5 \) suggests that even a basic smoothing operation on the noisy data can significantly enhance measurement statistics. As \( I_{\mathrm{Exp}} \) progressively converges to \( I_{\mathrm{GT}} \) with increasing detector counts beyond \( 10^6 \), \( D_\mathrm{Ref} \) continues to exhibit a plateau, indicating residual distortions introduced by the smoothing procedure. However, the consistent observation that \( D_{\mathrm{GPR}} \leq D_{\mathrm{Ref}} \) and \( D_{\mathrm{GPR}} \leq D_{\mathrm{Exp}} \) for any detector count underscores the effectiveness of GPR in refining the posterior distribution beyond what is achievable through smoothing alone.

Since both \( D_{\mathrm{GPR}} \) and \( D_{\mathrm{Exp}} \) decrease monotonically with increasing time, one can define the detector count \( N_{\mathrm{GPR}} \) required when using GPR to achieve the same data quality that would otherwise require \( N_{\mathrm{Exp}} \) as:  
\begin{equation}
    N_{\mathrm{GPR}} = D_{\mathrm{GPR}}^{-1}(D_{\mathrm{Exp}}(N_{\mathrm{Exp}})).
\end{equation}  
Fig.~\ref{fig:6}(b) presents the ratio \( \frac{N_{\mathrm{Exp}}}{N_{\mathrm{GPR}}} \) as a function of relative entropy, measured in nats. For this set of synthetic scattering intensities from the interacting hard sphere system, it is evident that the neutron flux required to achieve the same accuracy using GPR is approximately an order of magnitude lower than that required without GPR for relative entropy values greater than 2 nats. In this region, the GPR approach effectively accelerates data acquisition and enhances data quality by mitigating random noise from the experimental process. However, as accuracy improves below 2 nats, the acceleration provided by GPR becomes less significant, since both the GPR predictions and experimental data rapidly converge to the ground truth.

\section{Experimental Validations}
\label{sec:results}

In this section, we present the experimental validation of several commonly encountered soft matter systems to assess the feasibility and efficacy of the GPR-based statistical inference for SANS measurements. All SANS measurements were conducted at the Extended Q-range small-angle neutron scattering diffractometer (EQSANS) at the Spallation Neutron Source (SNS) at Oak Ridge National Laboratory (ORNL). The measured scattering intensities were corrected for detector background, sensitivity, and empty cell scattering. Additionally, the intensities were normalized to absolute units using a porous silica standard sample \cite{Zhao2010a, Heller2018}. 

Due to its time-of-flight nature, the scattering intensities can be reduced at any arbitrary exposure time, provided it falls within the overall measurement duration. This inherent property is crucial, as it enables a rigorous and precise quantitative evaluation of statistical inference enhancements, ensuring robust and reliable analysis of SANS data.

\subsection{Experimental Validation of GPR-Based Inference Using SDS Micellar Solutions}

The first example examined is the SANS spectrum of an aqueous solution of sodium dodecyl sulfate (SDS). A solution of SDS and lithium chloride (LiCl) was prepared with a fixed SDS concentration of 50 mg/mL and a LiCl:SDS molar ratio of 1 \cite{SDS}. The samples were prepared by accurately weighing the required amounts of SDS surfactant and LiCl salt, followed by the addition of the appropriate volume of water to achieve the target concentrations. 

For this specific measurement, a sample-to-detector distance of 1.3 m and an incident neutron wavelength of 6 \AA\ were selected, covering a $Q$ range of 0.01 to 0.3 \AA$^{-1}$. The SDS micellar sample was housed in a banjo cell with a path length of 1 mm, and all measurements were conducted at 25 $^{\circ}$C.

\begin{figure}[h]
    \centering
    \includegraphics[width=1.0\textwidth]{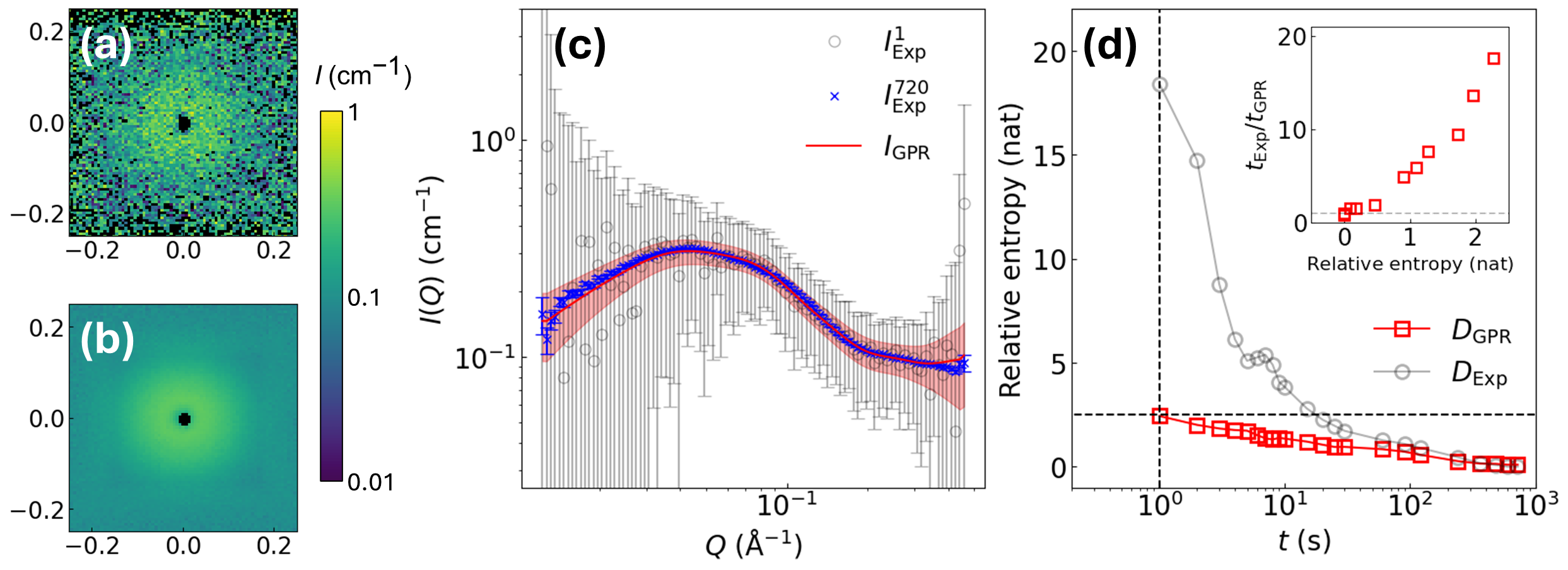}  
    \caption{(a, b) Two-dimensional neutron scattering intensity distributions of an aqueous sodium dodecyl sulfate (SDS) micellar solution measured using the Extended Q-range Small-Angle Neutron Scattering (EQSANS) instrument at the Spallation Neutron Source (SNS) with exposure times of (a) 1 s and (b) 720 s. (c) One-dimensional scattering intensity profiles comparing the low-count measurement ($I_{\text{Exp}}^{1s}$, circles) with the high-count measurement ($I_{\text{Exp}}^{720s}$, crosses), demonstrating statistical noise reduction with increased acquisition time. The Gaussian Process Regression (GPR)-based inference ($I_{\text{GPR}}$) is shown as a red curve, with the shaded region representing the inferred confidence interval. (d) The evolution of relative entropy ($D_{\text{GPR}}$ and $D_{\text{Exp}}$) as a function of measurement time, illustrating the efficiency of GPR in reducing statistical uncertainties. The inset shows the ratio $t_{\text{Exp}}/t_{\text{GPR}}$, highlighting a 20-fold reduction in measurement time required to achieve comparable statistical accuracy through GPR inference.}

    \label{fig:7}
\end{figure}

Figs.~\ref{fig:7}(a) and (b) illustrate the two-dimensional distribution of detector counts for scattered neutrons, corresponding to measurement times of 1 second and 720 seconds, respectively. As indicated by the color bar representing the scaled scattering intensity, a distinct contrast emerges between these two cases. At shorter counting times, the distribution appears sparse and exhibits pronounced statistical fluctuations due to the limited number of detected neutrons. 

Fig.~\ref{fig:7}(c) presents the one-dimensional intensity profiles with an exposure time of 1 s ($I_{\mathrm{Exp}}^1$, circles) and 720 s ($I_{\mathrm{Exp}}^{720}$, crosses). These experimental measurements are compared with the scattering intensity inferred from the GPR model, denoted as $I_{\mathrm{GPR}}$. In this representation, the red curve corresponds to the inferred mean value, while the red-shaded region indicates the confidence interval within one standard deviation.

In this analysis, $I_{\mathrm{Exp}}^{720}$ serves as the reference spectrum, considered as the ground truth in prior computational benchmarking studies, despite the presence of inherent noise. The intensity fluctuations exhibit a magnitude of approximately 1\% of the mean at each sampled $Q$ point. As expected, a direct comparison between $I_{\mathrm{Exp}}^1$ and $I_{\mathrm{Exp}}^{720}$ demonstrates that the enhancement in counting statistics over an extended measurement period effectively suppresses erratic fluctuations in the mean scattering intensity across the probed $Q$ range. Moreover, the prolonged acquisition time significantly mitigates large experimental uncertainties, yielding a more consistent and reliable representation of the scattering intensity.

Furthermore, the mean value of \( I_{\mathrm{GPR}} \), inferred from \( I_{\mathrm{Exp}}^1 \), shows improved quantitative agreement with \( I_{\mathrm{Exp}}^{720} \), as evidenced by the suppression of random noise.
Notably, the confidence interval associated with \( I_{\mathrm{GPR}} \) is significantly smaller than the experimental uncertainties of \( I_{\mathrm{Exp}}^1 \) of the magnitude of random noise, owing to the correlations between neighboring bins. The reductions vary by orders of magnitude depending on the bin size and experimental uncertainties, as described by Eqn.~\eqref{eq:10}.

Fig.~\ref{fig:7}(d) illustrates the comparison between the two $Q$-averaged relative entropies, $D_\mathrm{GPR}$ (squares) and $D_\mathrm{Exp}$ (circles), as functions of the experimental measuring time. Here, the measuring time is proportional to the scattered neutron counts, as shown on the x-axis in Fig.~\ref{fig:6}. First, we discuss the temporal evolution of $D_\mathrm{Exp}$. As the counting time increases, a more pronounced decrease is observed compared to the trend presented in Fig.~\ref{fig:6}(b). It is crucial to emphasize that, in our synthetic data used for computational benchmarking, the only source of noise considered is counting statistics, which scales with the square root of the counts. Other potential noise sources---such as background noise and electronic noise \cite{knoll2010radiation, lannunziata2012handbook}---that could introduce significant errors, particularly at low count rates, are not accounted for in this framework. We hypothesize that these unaccounted factors also influence the steeper decrease in $D_\mathrm{Exp}$. Their impact depends on the specifics of the experiment and the nature of the measurements performed.

When the counting time is shorter than 5 s, $D_\mathrm{GPR}$ exhibits a significant reduction relative to $D_\mathrm{Exp}$ due to noise suppression. As the counting time exceeds 50 seconds, $D_\mathrm{Exp}$ decreases to approximately 1 and gradually converges toward $D_\mathrm{GPR}$, though it remains slightly larger than $D_\mathrm{GPR}$. This trend indicates diminishing informational returns in applying GPR inference for $I_{\mathrm{Exp}}$ as counting time increases. Furthermore, $D_\mathrm{GPR}$ steadily decreases with counting time. After approximately 400 seconds, it becomes indistinguishable from $D_\mathrm{Exp}$, with both values approaching zero. This behavior demonstrates that GPR inference progressively improves over time, ultimately yielding results nearly identical to the ground truth after sufficient data acquisition. 

The vertical dashed line in Fig.~\ref{fig:7}(d) shows that GPR-based inference reduces the relative entropy of a 1-second measurement from 18 nats to approximately 2 nats, while the horizontal dashed line indicates that achieving this level of precision experimentally would require approximately 20 seconds. Consequently, GPR accelerates data acquisition by a factor of 20, signifying a 20-fold enhancement in counting efficiency. This implies that GPR significantly reduces the required measurement time while maintaining, or even improving, the accuracy of inferred scattering intensity.

To quantify the enhancement in counting efficiency, we analyze the ratio $\frac{t_{\mathrm{Exp}}}{t_{\mathrm{GPR}}}$ as a function of relative entropy as defined in Eq.~\eqref{eq:21}. The inset of Fig.~\ref{fig:7}(d) illustrates these results. The horizontal line at $\frac{t_{\mathrm{Exp}}}{t_{\mathrm{GPR}}} = 1$ represents the limit where GPR achieves statistical accuracy equivalent to SANS experimental observations for the same measurement duration. This visualization provides a clear perspective on the efficacy of GPR-based statistical inference in improving data acquisition efficiency.  

Across the entire 720-second experimental range, the statistical enhancement varies; however, the inferred counting efficiency consistently surpasses that of direct experimental measurements. This underscores the robust advantage of GPR in suppressing statistical uncertainties and refining scattering intensity estimation, enabling more efficient data acquisition without compromising accuracy.

\subsection{Comparative Analysis of Measurement Efficiency Across Soft Matter Systems}

Further validation and analysis of the existing EQSANS data for various soft matter systems, along with analytical scattering functions for different soft materials, are provided in Appendix~\ref{app:b}. Figure~\ref{fig:8}(a) presents a summary of the results, displaying a scatter plot that compares the experimental measurement time, \( t_\mathrm{Exp} \), with the GPR-accelerated measurement time, \( t_\mathrm{GPR} \). The plot includes data from all EQSANS experiments (represented by open symbols) and computational benchmarks (represented by filled symbols) across various classes of soft materials.

\begin{figure}[h]
    \centering
    \includegraphics[width=\textwidth]{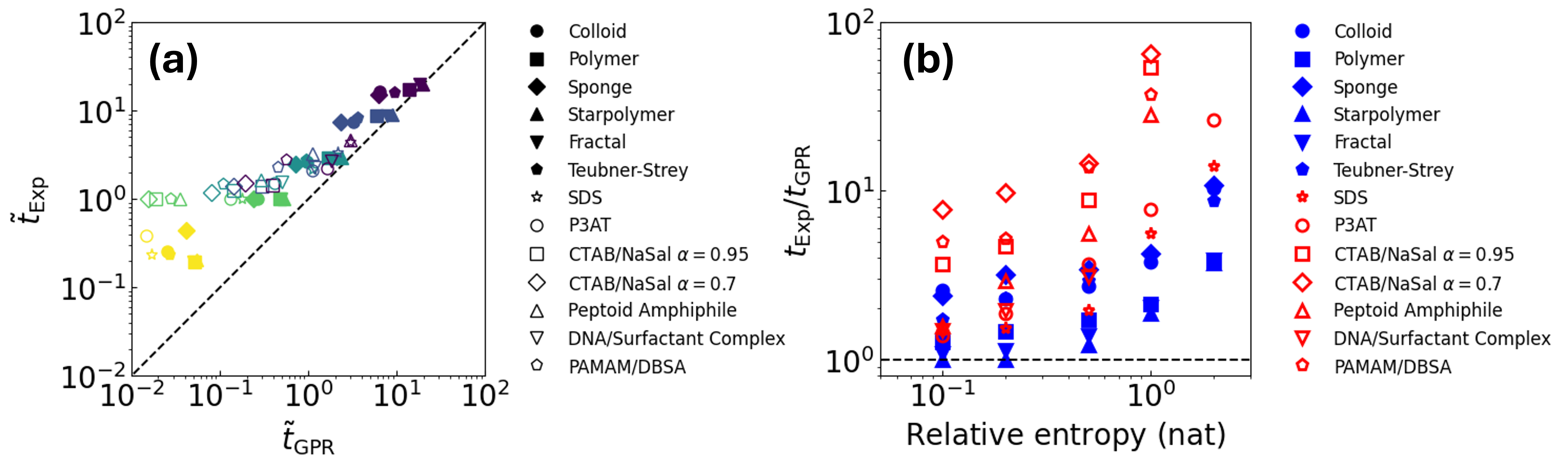}  
\caption{(a) Comparison of scaled experimental measurement time (\(\tilde{t}_\mathrm{Exp}\)) and GPR-inferred scaled time (\(\tilde{t}_\mathrm{GPR}\)) for various soft matter systems. The dashed 45-degree line represents the scenario where GPR inference provides no enhancement. All tested data points lie above this line, confirming that GPR-based statistical inference significantly reduces measurement time while preserving the statistical accuracy of SANS spectra. This demonstrates that GPR effectively enhances counting efficiency, optimizing experimental efforts without compromising data quality. (b) Measured acceleration factor (\(t_\mathrm{Exp}/t_\mathrm{GPR}\)) as a function of relative entropy for different soft matter systems. The dashed horizontal line at \(t_\mathrm{Exp}/t_\mathrm{GPR} = 1\) indicates no improvement in measurement time. The results reveal that for relative entropy values below 0.1, the acceleration factor remains close to unity, suggesting minimal enhancement due to the already high statistical quality of the data. However, for relative entropy values between 0.1 and 1, the acceleration factor increases substantially, often reaching values up to 10. At even higher relative entropy values (\(>1\)), the acceleration factor can exceed 10, with some cases showing reductions in measurement time by nearly two orders of magnitude. This highlights the robustness of GPR in significantly improving measurement efficiency, particularly for low-signal or high-noise conditions. Additionally, real experimental measurements (red open symbols) exhibit a more pronounced enhancement compared to synthetic data, further emphasizing the effectiveness of GPR in practical experimental scenarios.}
    \label{fig:8}
\end{figure}

To facilitate a direct comparison of counting efficiency improvements across different sample spectra, the measurement time is expressed in terms of a normalized scaled time, where both the experimentally measured and GPR-inferred times are normalized by the measurement duration required to achieve a statistical accuracy of unity. For synthetic data, a similar normalization is applied using the total detector count required for the same accuracy threshold.

The dashed 45-degree line in Fig.~\ref{fig:8}(a) represents the scenario where the GPR-inferred scaled time matches the experimentally measured scaled time, signifying no improvement in counting efficiency through GPR inference. Notably, for the deep purple markers corresponding to a very low relative entropy of 0.1, the GPR-inferred spectra preserve nearly identical structural information as the direct experimental measurements, indicating that the random noise in the experimental data has been sufficiently suppressed in long-duration measurements.

A key observation, as evidenced by the color evolution of symbols representing the value of \( D \), is that all tested data points—both from synthetic spectra (filled symbols) and experimental EQSANS measurements (open symbols)—consistently lie above the dashed line. This strongly confirms the effectiveness of GPR-based statistical inference in significantly reducing the required measurement time while preserving the statistical accuracy of the SANS spectra. 

The acceleration factor, defined as \( t_\mathrm{Exp} / t_\mathrm{GPR} \), quantifies this efficiency gain as a function of relative entropy to demonstrate its dependence on data quality. The results are displayed in Fig.~\ref{fig:8}(b). Here, the results from the synthetic data and experimental measurements are respectively represented by blue filled symbols and red open symbols. The data spans nearly two orders of magnitude across various examined systems.

For relative entropy values below 0.1, the acceleration factor remains near unity, indicating negligible improvement due to the already high statistical quality of the experimental data. However, for systems with relative entropy between 0.1 and 1, the acceleration factor increases substantially, often reaching values up to 10. This indicates that GPR can reduce measurement time by an order of magnitude while preserving spectral accuracy.

For higher relative entropy exceeding 1, the acceleration factor can surpass 10, with certain data points achieving reductions in measurement time by nearly two orders of magnitude. This remarkable efficiency gain demonstrates that GPR is particularly beneficial for low-signal or high-noise conditions, where traditional experimental averaging would require significantly longer acquisition times for gaining the data of same quality. 

Additionally, different soft matter systems exhibit varying degrees of acceleration, with SDS and P3AT showing the most substantial improvements, while colloidal and polymeric systems tend to have more moderate gains. This variability likely arises from differences in intrinsic scattering contrast and statistical noise levels across different materials. Importantly, the absence of data points below the dashed line at an acceleration factor of one further underscores the robustness of GPR in efficiently predicting spectral information across diverse experimental conditions.

Moreover, the enhancement of real experimental measurements (red open symbols) is observed to be significantly higher than that of the synthetic data, roughly by an order of magnitude. We attribute this observation to various sources of statistical noise that are not considered in the synthetic data, which accounts only for counting statistics.

Overall, these results confirm that GPR effectively enhances counting efficiency by leveraging statistical correlations within the data to infer reliable spectral intensities with reduced experimental effort. Consequently, the application of GPR not only optimizes measurement time but also facilitates high-throughput SANS experiments without compromising data quality.

\section{Conclusions}
\label{sec:conclusion}

We have developed a Bayesian statistical inference approach based on Gaussian Process Regression (GPR) to enhance the robustness of small-angle neutron scattering (SANS) data obtained from measurements with a low signal-to-noise ratio (SNR). Unlike conventional machine-learning methods that require extensive training datasets, our method employs a one-shot inference strategy that directly reconstructs high-fidelity SANS data by leveraging the intrinsic smoothness and continuity of the scattering function. This capability effectively suppresses noise without necessitating large-scale data collection or prior instrument-specific training.

The effectiveness of this approach has been validated through computational benchmarks and experimental SANS measurements at the Extended Q-range Small-Angle Neutron Scattering (EQSANS) instrument at the Spallation Neutron Source (SNS). Experimental results demonstrate that our \textit{one-shot} inference method improves measurement efficiency by one to two orders of magnitude, significantly reducing acquisition times while maintaining accuracy comparable to considerably longer measurements.

This substantial gain in measurement efficiency opens new possibilities for SANS experiments, enabling more effective and versatile data collection across a wide range of applications:

\begin{itemize}
    \item \textbf{Enhanced experimental efficiency:} By enabling more sample measurements without hardware modifications, our approach optimizes neutron beamtime usage and increases experimental throughput while preserving data quality. The reduction in required acquisition time facilitates high-throughput studies, making SANS more accessible across multiple disciplines.
    \item \textbf{Advancements in kinetic and in situ studies:} The method is particularly beneficial for studies involving rare or difficult-to-synthesize materials, where limited sample availability often constrains experimental feasibility.
    \item \textbf{Advancements in kinetic and in situ studies:} Our approach enables precise monitoring of mechanically driven systems and irreversible transformations occurring on timescales shorter than conventional measurement durations, significantly improving real-time structural investigations.
    \item \textbf{Improved contrast variation studies:} By improving statistical robustness, the framework addresses challenges associated with weak cross-correlation signals in multicomponent systems where isotopic substitution reduces scattering contrast.
    \item \textbf{Broader applicability beyond SANS:} The proposed framework extends to various experimental techniques that suffer from low SNR. For example, it can be directly applied to structural studies using laboratory-based small-angle x-ray scattering (SAXS)\cite{Pedersen2025}, reducing reliance on synchrotron radiation sources while maintaining high-fidelity data reconstruction.
    \item \textbf{Advancements in dynamical studies:} The method enhances neutron spin echo (NSE)\cite{JCNS}, quasielastic neutron scattering (QENS) \cite{Boutin1968, Plazanet2025}, and inelastic x-ray scattering (IXS) \cite{Baron2016} by improving statistical robustness and optimizing measurement times, advancing the study of relaxation dynamics, diffusion, and molecular motion.
    \item \textbf{Potential applications in compact neutron sources (CANS):} One of the most promising applications of our approach is in SANS at CANS \cite{Bruckel2020}. While CANS operate at inherently lower neutron flux than large-scale facilities, our framework significantly enhances data utilization by extracting maximum information from limited scattering signals. This improvement opens new possibilities for high-resolution structural analysis, kinetic process characterization, and time-resolved studies, making advanced neutron experiments more feasible in resource-constrained environments.
\end{itemize}

By eliminating the dependence on pre-trained models and large datasets, our one-shot statistical inference framework represents a fundamental shift in neutron and x-ray scattering methodologies. Future work will focus on optimizing computational efficiency for real-time data analysis and refining parameter selection for diverse experimental conditions. By integrating advanced statistical inference with next-generation scattering methodologies, this approach has the potential to transform neutron and x-ray scattering workflows, enabling groundbreaking discoveries across physics, chemistry, and biology.

\begin{acknowledgments}

This research at ORNL's Spallation Neutron Source was sponsored by the Scientific User Facilities Division, Office of Basic Energy Sciences, U.S. Department of Energy. It was also supported by the Laboratory Directed Research and Development Program of Oak Ridge National Laboratory, managed by UT-Battelle, LLC, for the U.S. Department of Energy. Beam time was allocated to EQSANS under proposal numbers IPTS-22170.1, 22386.1, 23463.1 and 25953.1. Y.S. was supported by the U.S. Department of Energy, Office of Science, Office of Basic Energy Sciences, Materials Sciences and Engineering Division, under Contract No. DE-AC05-00OR22725.

\end{acknowledgments}

\appendix

\section{Incorporating Instrumental Resolution into Probabilistic Inference}
\label{app:a}

It is important to note that the scattering spectra obtained in SANS experiments are inherently influenced by instrumental resolution \cite{Pedersen2025}, which can be affected by a combination of factors. Assuming the uncertainty from the instrument is assigned such that 
\[
Q_i \sim \mathcal{N}(\mu_{Q,i}, \sigma_{Q,i}^2),
\]
the separation between $Q_i$ and $Q_j$ can be evaluated as
\[
\Delta Q_{ij} = Q_j - Q_i,
\]
and it should follow the distribution
\[
\Delta Q_{ij}\sim \mathcal{N}(\mu_{Q,j} - \mu_{Q,i}, \sigma_{Q,j}^2 + \sigma_{Q,i}^2).
\]

Let $\Delta\mu_{ij} \equiv \mu_{Q,j} - \mu_{Q,i}$ and $\Sigma_{ij}^2 \equiv \sigma_{Q,j}^2 + \sigma_{Q,i}^2$. The correlation between $I(Q_i)$ and $I(Q_j)$ should be assigned as
\[
k_{ij} = \frac{\int_{-\infty}^\infty \exp{\left(-\frac{\|z\|^2}{2\lambda^2}\right)} \exp\left(-\frac{\|\Delta\mu_{ij}-z\|^2}{2\Sigma_{ij}^2}\right) \, d z}{\int_{-\infty}^\infty \exp\left(-\frac{\|\Delta\mu_{ij}-z\|^2}{2\Sigma_{ij}^2}\right) \, d z}.
\]

\[
=\sqrt{\frac{\lambda^2}{\lambda^2 + \Sigma_{ij}^2}} \exp{\left(-\frac{\|\Delta\mu_{ij}\|^2}{2(\lambda^2 + \Sigma_{ij}^2)}\right)}.
\label{eq:k_resolution}
\]

The GPR posterior considering instrument resolution can thus be obtained by reformatting Eqn~\ref{eq:4} accordingly. Note that when $\lambda^2 \gg \sigma_j^2 + \sigma_i^2$, the result simplifies to the case without accounting for instrumental resolution. In contrast, if $\lambda^2 \ll \sigma_j^2 + \sigma_i^2$, then $k_{ij} \to 0$, and the correlation between $I(Q_i)$ and $I(Q_j)$ vanishes.

\section{Further Experimental Validations and Analysis}
\label{app:b}

\subsection{Computationally Generated Synthetic Data}

\begin{figure}[h]
    \centering
    \includegraphics[width=\textwidth]{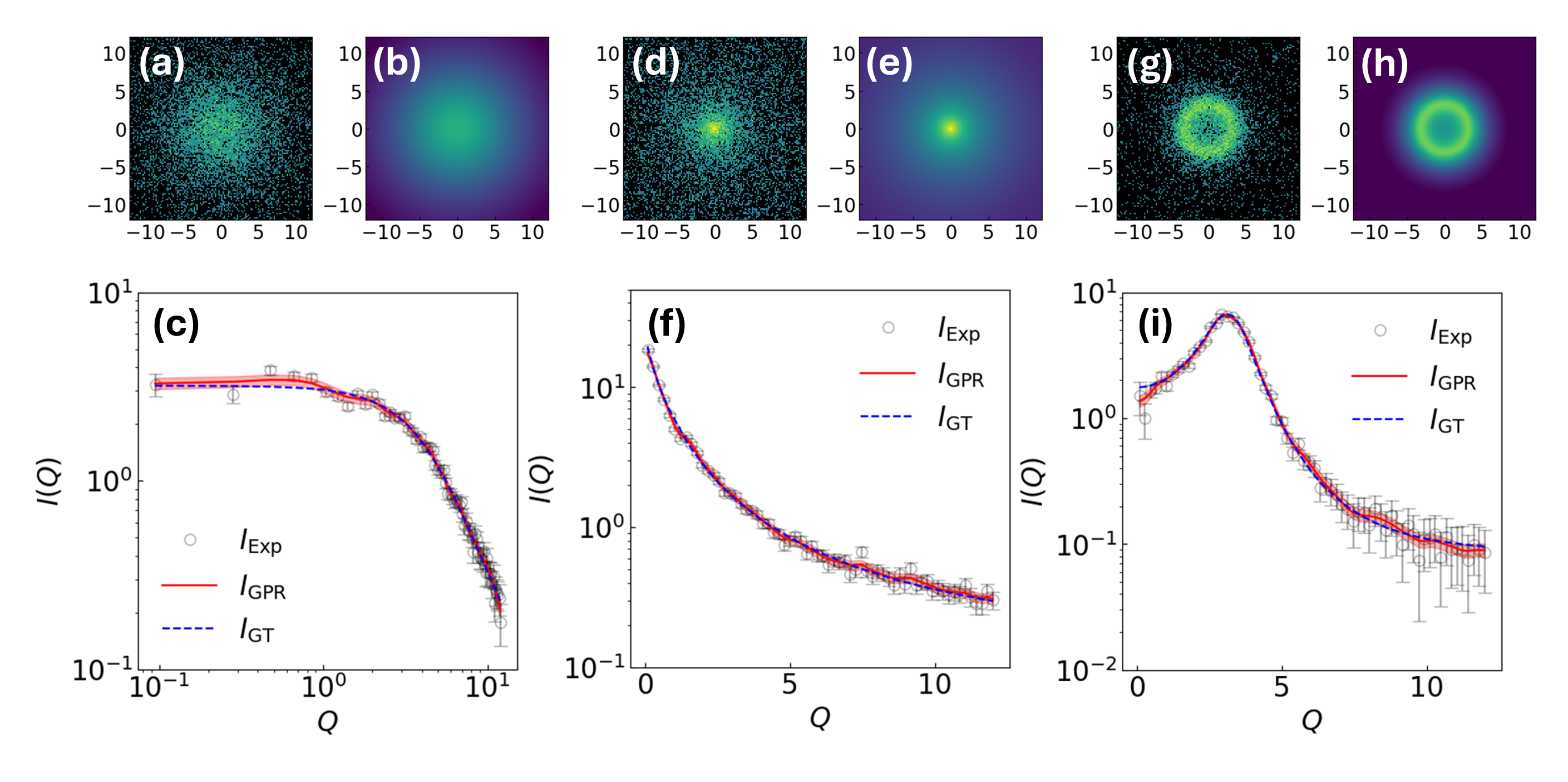}  
\caption{(a, b) Two-dimensional computationally synthesized scattering spectra of a star polymer \cite{benoit1953effect}, calculated with total detector counts of $10^{4}$ and infinite counts, respectively.  
(c) Azimuthally averaged one-dimensional intensity profiles: the dashed line represents the theoretical ground truth intensity $I_\mathrm{GT}$; circles denote the noisy experimental intensity $I_\mathrm{Exp}$ from (a); the shaded region indicates the GPR-predicted confidence interval $I_\mathrm{GPR}$; and the solid line represents the mean of $I_\mathrm{GPR}$.  
(d, e) Scattering spectra of a pearl necklace model with a fractal dimension characteristic of micellar solutions \cite{Chen1986}, computed under the same detector count conditions as (a, b).  
(f) Comparison of the $I_\mathrm{GPR}$ confidence interval with uncertainties in $I_\mathrm{Exp}$, demonstrating agreement between the mean of $I_\mathrm{GPR}$ and $I_\mathrm{GT}$ over the probed $Q$ range.  
(g, h) Two-dimensional scattering spectra of the Teubner-Strey model of microemulsions \cite{Teubner1987}.  
(i) Validation of GPR-based statistical inference for microemulsions.}  
    \label{fig:b1}
\end{figure}

Fig.~\ref{fig:b1} presents a comprehensive analysis of computationally synthesized scattering spectra for different models, illustrating the impact of detector count conditions and the effectiveness of GPR-based statistical inference. Panels (a) and (b) show two-dimensional computationally synthesized scattering spectra of a star polymer \cite{benoit1953effect}, calculated with total detector counts of $10^{4}$ and infinite counts, respectively. The difference between these spectra highlights the effect of detector noise on the observed scattering patterns. Panel (c) presents the azimuthally averaged one-dimensional intensity profiles. The dashed line represents the theoretical ground truth intensity, $I_\mathrm{GT}$, while the circles correspond to the noisy experimental intensity, $I_\mathrm{Exp}$, obtained from panel (a). The shaded region indicates the confidence interval predicted by GPR, $I_\mathrm{GPR}$, and the solid line represents the mean of $I_\mathrm{GPR}$. This comparison demonstrates the ability of GPR to reconstruct the underlying scattering intensity while quantifying uncertainties.  

Panels (d) and (e) depict scattering spectra of a pearl necklace model, which exhibits a fractal dimension characteristic of micellar solutions \cite{Chen1986}. These spectra are computed under the same detector count conditions as in panels (a) and (b), allowing for a direct comparison of the effect of noise on different structural models. Panel (f) further investigates the agreement between the GPR-predicted intensity and experimental data. The $I_\mathrm{GPR}$ confidence interval is compared with the uncertainties in $I_\mathrm{Exp}$, demonstrating that the mean of $I_\mathrm{GPR}$ closely follows $I_\mathrm{GT}$ across the probed $Q$ range.  

Panels (g) and (h) illustrate two-dimensional scattering spectra of the Teubner-Strey model, which describes microemulsions \cite{Teubner1987}. Finally, panel (i) validates the application of GPR-based statistical inference for microemulsions, confirming its robustness in analyzing complex scattering data.  

Overall, this figure highlights the effectiveness of GPR in reconstructing scattering intensity profiles while providing well-calibrated uncertainty estimates across different models and experimental conditions.

\subsection{Experimental Data from EQSANS at SNS}

\begin{figure}[h]
    \centering
    \includegraphics[width=\textwidth]{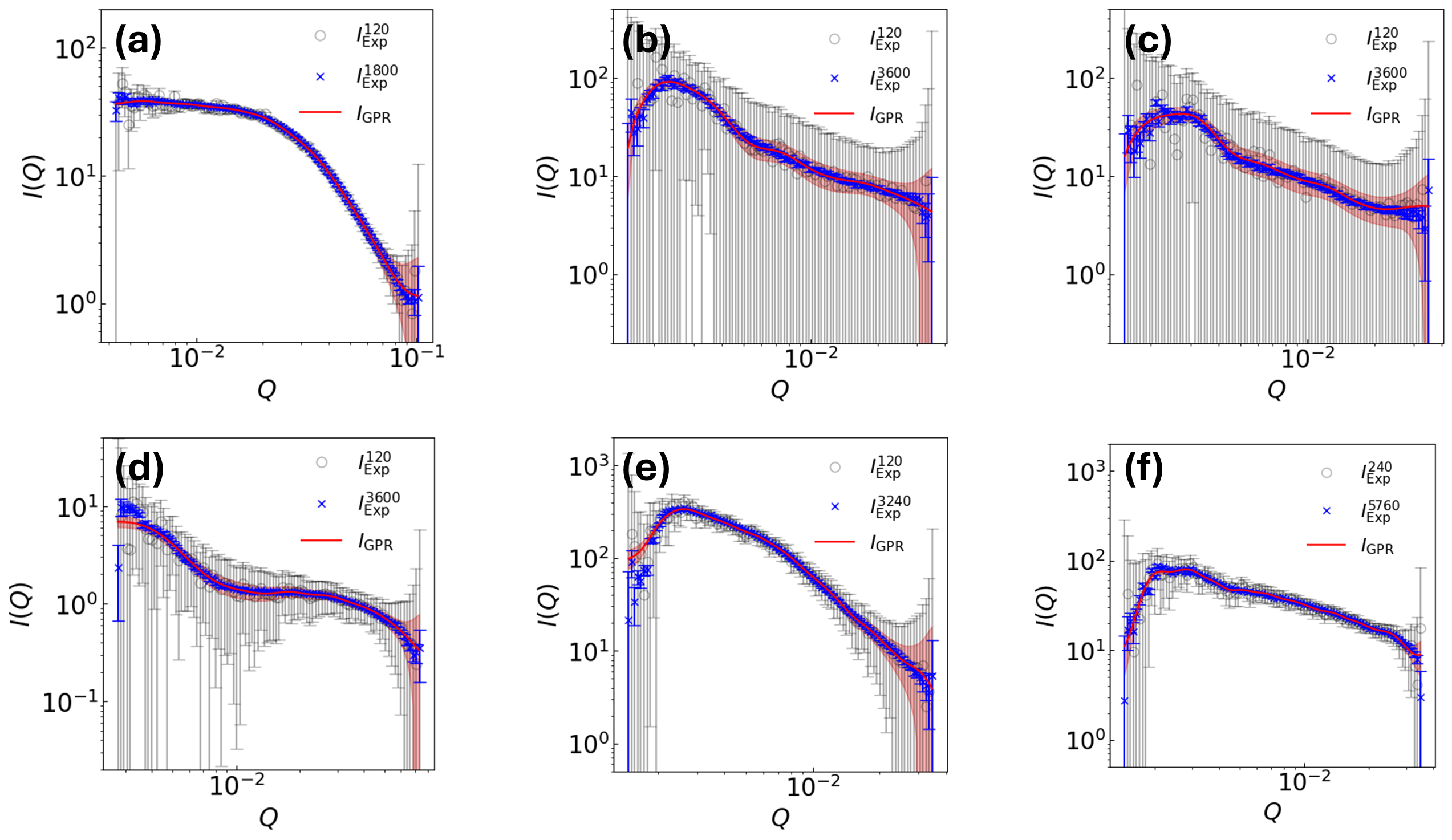}  
\caption{SANS measurements at EQSANS for various soft matter systems.  
(a) Solution of conjugated polymers with poly(3-alkylthiophene) (P3AT) backbones and alkyl side chains.  
(b, c) Aqueous CTAB/NaSal solutions with contrast variation (\(\alpha = 1, 0.7\)), where \(\alpha\) represents the atomic fraction of \(\mathrm{D_2O}\) in the mixed solvent system (\(\mathrm{H_2O} + \mathrm{D_2O}\)).  
(d) \(\mathrm{D_2O}\) solution of a peptoid amphiphile.  
(e) \(\mathrm{D_2O}\) solution of a DNA complex with a pH-sensitive gemini surfactant. 
(f) \(\mathrm{D_2O}\) solution of the self-assembly of a PAMAM dendrimer with dodecylbenzene sulfonic acid (DBSA).  
Gray circles and blue crosses represent experimental data at different exposure times, while the red curve and shaded region denote the GPR inference and uncertainty. The results highlight the effectiveness of statistical inference in enhancing data quality and correcting experimental artifacts.}
    \label{fig:b2}
\end{figure}

Fig.~\ref{fig:b2} presents a series of SANS measurements performed at EQSANS, highlighting various soft matter systems. These include: (a) a solution of conjugated polymers with poly(3-alkylthiophene) (P3AT) backbones and alkyl side chains; (b) an aqueous solution of cetyltrimethylammonium bromide (CTAB) in 0.95 M sodium salicylate (NaSal) \cite{Lam2019CTABNaSal} with \(\alpha = 1\); (c) an aqueous CTAB/NaSal solution with \(\alpha = 0.7\), where \(\alpha\) represents the atomic fraction of \(\mathrm{D_2O}\) in the mixed solvent system (\(\mathrm{H_2O} + \mathrm{D_2O}\)); (d) a \(\mathrm{D_2O}\) solution of a peptoid amphiphile; (e) a \(\mathrm{D_2O}\) solution of a DNA complex with a pH-sensitive gemini surfactant; and (f) a \(\mathrm{D_2O}\) solution of the self-assembly of a PAMAM dendrimer with dodecylbenzene sulfonic acid (DBSA). The superscript of \(I_\mathrm{Exp}\) denotes the experimental measurement time at EQSANS.

Across all tested systems, our statistical inference approach demonstrates a significant enhancement in data quality, underscoring its robustness and efficiency. The results in panels (b) and (c) highlight the role of contrast variation, a widely employed technique in SANS to investigate structural heterogeneity in soft materials. A known challenge in this approach is the degradation of statistical quality due to reduced coherent scattering from isotopic labeling. However, our results show that statistical inference can effectively mitigate this limitation, thereby enhancing the feasibility of contrast variation SANS for material characterization.


%

\end{document}